\documentclass[prl,aps,preprintnumbers,nofootinbib,notitlepage]{revtex4-1}
\usepackage{epsfig,graphics}
\usepackage{amsfonts,amssymb,amsmath}            
\usepackage{ifthen}
\usepackage{amsthm}
\usepackage{hyperref}
\usepackage{url}
\usepackage{todonotes}
\usepackage{color, colortbl,hhline}
\usepackage{tikz}
\usetikzlibrary{matrix}
\usepackage{diagbox}

\newcommand{\comment}[1]{}

\theoremstyle{plain}

\theoremstyle{definition}

\definecolor{amber}{rgb}{1.0, 0.75, 0.0}
\definecolor{aureolin}{rgb}{0.99, 0.93, 0.0}
\definecolor{steel}{HTML}{91CDE2}
\definecolor{bblue}{HTML}{508AA8}
\definecolor{beyes}{HTML}{EDF6FC}

\begin{document}

\title{The Binary-Outcome Detection Loophole}
\date{\today}

\author{Thomas \surname{Cope}}
\email{thomas.cope@itp.uni-hannover.de}
\address{Institut f{\'u}r Theoretische Physik, Leibniz Universit{\'a}t Hannover, Appelstr. 2, 30167 Hannover, Germany}

\begin{abstract}
The detection loophole problem arises when quantum devices fail to provide an output for some runs. If treating these devices in a device-independent manner, failure to include the unsuccessful runs in the output statistics can lead to an adversary falsifying security i.e. Bell inequality violation. If the devices fail with too high frequency, known as the \emph{detection threshold}, then no security is possible, as the full statistics cannot violate a Bell inequality.\\
In this work we provide an intuitive local hidden-variable strategy that the devices may use to falsify any two-party, binary-outcome no-signalling distribution up to a threshold of $2(m_{\mathrm{A}}+m_{\mathrm{B}}-8)/(m_{\mathrm{A}}m_{\mathrm{B}}-16)$, where $m_{\mathrm{A}}$, $m_{\mathrm{B}}$ refer to the number of available inputs choices to the two parties. This value is the largest analytically predicted lower bound for no-signalling distributions. We strongly conjecture it gives the true detection threshold for $m_{\mathrm{A}}=m_{\mathrm{B}}$, and for computationally tractable scenarios we provide the Bell inequality which verifies this. We also prove that a non-trivial detection threshold remains, even when allowing one party an arbitrary number of input choices.\\

\end{abstract}
\maketitle

\section{Introduction}
Due to the scales on which it operates, quantum technology faces the challenge of single photons or electrons being lost to the environment. This can result in devices failing to give any output. Ignoring these failures leads to the `detection loophole' \cite{P1970,CH1974,F1982} security flaw. This is where a preprogrammed `hidden-variable' device can falsely appear to exhibit non-local behaviour. Non-locality is necessary for the security proofs of device-independent quantum cryptography \cite{MY1998,BHK,AGM,VV2,ColbeckThesis,PAMBMMOHLMM,CK2,MS1}, therefore understanding and preventing the detection loophole is an extremely relevant problem.\\
 One important question to consider  is how low the rate of successful detection events (the efficiency) can be before all observed correlations are describable by a local realistic model.  Knowing this threshold allows one to set minimum requirements for commercial devices and benchmark current technology. However, obtaining this bound for quantum states is generally difficult due to the infinite set of extremal quantum correlations, and only a few optimal constructions are known \cite{E1993,VPB2010}.\\
In this article we present an intuitive local hidden-variable (LHV) construction for two parties, arbitrary inputs, and binary outputs, which will be able to reproduce any no-signalling distribution obtained by the successful runs, up to a detection efficiency dependent on the number of inputs. This provides a lower bound on the threshold for quantum measurements in the same scenario. When both parties have the same number of inputs into their device, this construction achieves numerically known thresholds (for general no-signalling distributions) leading us to conjecture it is optimal for this symmetric case. We furthermore show that in cases with an asymmetric number of measurements, increasing the number of Bob's measurements $m_{\mathrm{B}}$ above $2^{\lceil \log_2 m_{\mathrm{A}}\rceil }$ provides no additional power in verifying non-local correlations.

Bell's seminal theorem \cite{B1964} and its subsequent generalisations \cite{CHSH1969,F1981,BCPSW2014} give fundamental constraints on the correlations exhibited by any local realistic model; constraints that quantum theory can violate. These violations have been confirmed experimentally  \cite{Aspect81, Tittel1998}. Due to limitations on technology however, to show Bell violations they relied on a `fair-sampling' assumption; that the device failures were non-malicious and the successful detections were representative of the underlying system. In cryptographic protocols however, we cannot make that assumption, allowing an adversary (Eve) to pre-program the device to fail. It wasn't until much later that loophole-free violations, with no fair-sampling assumptions, were experimentally demonstrated \cite{Giustina&,Hensen&,Shalm&}. The difficulty involved in closing this loophole highlights the importance of obtaining the best theoretical thresholds possible, so that minimal technological developments are required to perform secure protocols.\\

\section{Preliminaries}

In this paper, we are working in the \emph{device-independence framework}. We assume that two parties (named Alice and Bob) have been distributed a joint system, on which they can make measurement choices, also referred to as inputs (labelled by $x$ for Alice, and $y$ for Bob) and receive outcomes (labelled $a$ for Alice and $b$ for Bob). We characterise the joint system only by the conditional probability distribution $p(ab|xy)$, making no assumptions about the underlying state or measurements made. This is known as a \emph{black box} description. However, we do assume that Alice and Bob can isolate their systems, also referred to here as \emph{devices}, from communicating with each other. This imposes the \emph{no-signalling conditions}
\begin{align}
	\sum_{a} p(ab|xy)&= \sum_{a} p(ab|xy')\;\; \forall {b,x,y,y'}, \\
	\sum_{b} p(ab|xy)&= \sum_{b} p(ab|x'y)\;\; \forall {a,x,x',y}.
\end{align}
When the number of inputs and outputs are finite, so that $x\in\{0\ldots m_A-1\}$, $y\in\{0\ldots m_B-1\}$, $a\in\{0\ldots n_A-1\}$, $b\in\{0\ldots n_B-1\}$, then we may express any no-signalling probability distribution via the vector
$\mathbf{p}:=[p(00|00)\ldots  p(n_A-1\;n_B-1|m_A-1\;m_B-1)]$. The set of such vectors forms a convex set with finitely many extremal points, known as the no-signalling polytope, $\mathcal{NS}$. This restriction is known as the $(m_{\mathrm{A}},m_{\mathrm{B}},n_{\mathrm{A}},n_{\mathrm{B}})$-scenario.\\

Within this set is a strict subset \cite{PR1994} of \emph{quantumly realisable} distributions, $\mathcal{Q}$. Unlike the full no-signalling space, $\mathcal{Q}$ has an infinite number of extremal points, making it more difficult to deal with computationally. Strictly contained within $\mathcal{Q}$ is the set of local distributions, $\mathcal{L}$. Any distribution $p(ab|xy)$ within $\mathcal{L}$ has a local hidden variable model of the form
$p(ab|xy)=\int_{\Lambda} d\lambda \rho(\lambda) p(a|x,\lambda)p(b|y,\lambda)$. These distributions may always be expressed as convex combinations of \emph{deterministic} distributions $p(ab|xy)=\delta_{a,a_x}\delta_{b,b_y}$, which are finite in number. Geometrically, this means the structure of $\mathcal{L}$ is also a \emph{polytope}. \\

$\mathcal{L}$ may be equivalently described by a set of \emph{Bell inequalities}, linear inequalities of the form $\sum_{a,b,x,y} s_{ab}^{xy}p(ab|xy) \leq  k$, where $p(ab|xy)$ is our input-conditional joint distribution \cite{T1993}. There is a finite set of \emph{facet} Bell inequalities; if all facets are satisfied by $p(ab|xy)$ it must have a local hidden-variable model i.e. it belongs to $\mathcal{L}$. Thus violation of a Bell inequality is used to prove the impossibility of a local hidden-variable model. We will also often denote a Bell inequality by a vector $\mathbf{s}=(s_{00}^{00}\ldots s_{n_A-1,n_B-1}^{m_A-1,m_B-1})$, though one must also state the sign and magnitude of the inequality.\\

The typical detection loophole model; and the one considered in this article, is one in which the devices fail to detect with equal probability independently of each other \cite{MP2003}. Whilst not completely general, it is how we would expect the device to behave if the failures were `honest'; if we see autocorrelations, or correlations between the joint failures; this is a clear signal of adversarial manipulation. The model considered here adds an extra output to both parties to alter the original distribution $p(ab|xy)$ in the following way:
\begin{align}
&p_{\eta}(ab|xy)=\eta^2p(ab|xy),\nonumber\\
&p_{\eta}(Fb|xy)=\eta(1-\eta)p(b|y),\nonumber\\
&p_{\eta}(aF|xy)=\eta(1-\eta)p(a|x),\nonumber\\
&p_{\eta}(FF|xy)=(1-\eta)^2.\label{Eta}
\end{align}
One can see this as a linear map $D_{\eta}: \mathbf{p} \rightarrow \mathbf{p}_{\eta}$, from the set of no-signalling distributions in the $(m_{\mathrm{A}},m_{\mathrm{B}},n_{\mathrm{A}},n_{\mathrm{B}})$-scenario to those in the $(m_{\mathrm{A}},m_{\mathrm{B}},n_{\mathrm{A}}+1,n_{\mathrm{B}}+1)$-scenario. The quantity we are interested in is the (quantum) \emph{critical detection efficiency}, $\eta_{\mathrm{c}}:=\inf \{\eta \mid \exists \mathbf{p}\in \mathcal{Q}, \mathbf{p}_{\eta} \not\in \mathcal{L}\}$, where $\mathcal{Q},\mathcal{L}$ are considered in the $(m_{\mathrm{A}},m_{\mathrm{B}},n_{\mathrm{A}},n_{\mathrm{B}})$-scenario and $(m_{\mathrm{A}},m_{\mathrm{B}},n_{\mathrm{A}}+1,n_{\mathrm{B}}+1)$-scenario respectively.\\


 To check the membership criterion $\mathbf{p}_{\eta}\in \mathcal{L}$, we can calculate the \emph{local weight}. This is defined for an arbitrary distribution $\mathbf{q}$ as:
\begin{equation}
\max_{w\in[0,1]} \mathbf{q}= w \mathbf{q}^{\mathcal{L}} + (1-w)\mathbf{q}'
\end{equation}
where $\mathbf{q}^{\mathcal{L}}$ is a local distribution and $\mathbf{q}'$ is a general no-signalling distribution. This linear program (see the appendix for details) gives $w=1$ iff $\mathbf{q}$ is local.\\

For a given $(m_{\mathrm{A}},m_{\mathrm{B}},n_{\mathrm{A}},n_{\mathrm{B}})$-scenario, we can use the linear weight to lower bound the critical detection threshold $\eta_{\mathrm{c}}$ in the following way. For every extremal no-signalling distribution $\mathbf{p}^{\mathrm{NS}}_{j}$, we can calculate the local weight of successive distributions $\mathbf{p}^{\mathrm{NS}}_{j,\eta}$ - allowing us (e.g. by the binary chop algorithm) to determine the detection threshold of that particular distribution, $\eta_{j}$. By doing this for all extremal points, we find that at $\eta^* = \min_{j} \eta_{j}$, the entire $(m_{\mathrm{A}},m_{\mathrm{B}},n_{\mathrm{A}},n_{\mathrm{B}})$ no signalling space is mapped into the $(m_{\mathrm{A}},m_{\mathrm{B}},n_{\mathrm{A}}+1,n_{\mathrm{B}}+1)$ local polytope. Thus, $\eta^*$ is necessarily a lower bound of $\eta_{\mathrm{c}}$. We will refer to $\eta^*$ as the \emph{no-signalling threshold}.\\
This bounding technique was performed in \cite{CC2019} on $m_{\mathrm{A}},m_{\mathrm{B}} \leq 6$ and $n_{\mathrm{A}}=n_{\mathrm{B}}=2$ for both parties, until the exponential growth in the number of extremal $\mathcal{NS}$ points became too large for numerical calculations.\\ Reproducing the table of thresholds from \cite{CC2019} in table \ref{Table1}, there are two patterns one observes immediately; that for $m_{\mathrm{A}}=m_{\mathrm{B}}=m$ the bound appears to match $4/(m+4)$, and that, if one fixes $m_{\mathrm{A}}$, the bound for $m_{\mathrm{B}}$ decreases with each additional output until $m_{\mathrm{B}}=2^{\lceil \log m_{\mathrm{A}} \rceil}$. In this article we prove that indeed the threshold for all $m_{\mathrm{A}}=m_{\mathrm{B}}=m$ is bounded below by $4/(m+4)$, and that it remains constant for all $m_{\mathrm{B}} \geq 2^{\lceil \log m_{\mathrm{A}} \rceil}$. Instead of doing this via numerical results, we construct an explicit local hidden variable model for all $\mathbf{p}_{\eta}$ up to this threshold value.\\

\begin{table}
\begin{center}
\begin{tabular}{|>{\columncolor{beyes}[2pt][1.9pt]}c|c>{\columncolor{beyes}[2pt][1.9pt]}cc>{\columncolor{beyes}[2pt][1.9pt]}cc}
\hline
\diagbox{$m_{\mathrm{A}}$}{$m_{\mathrm{B}}$}&2&3&4&5&6\\
\hline
2&2/3&2/3&2/3&2/3&2/3\\
3&&4/7&5/9&5/9&5/9\\
4&&&1/2&1/2&1/2\\
5&&&&4/9& * \\
\end{tabular}
\caption[Fundamental bounds on the detection threshold]{Cases for which the no-signalling threshold has been numerically calculated; these provide a lower bound on the corresponding critical detection efficiency for the quantum set. The * indicates numerical evaluation was not attained \cite{CC2019}.}
\label{Table1}
\end{center}
\end{table}


\section{Pre-Existing Local Hidden Variable Constructions}\label{PreStrat}
In order to understand our explicit construction, it is first useful to compare it to a local hidden-variable construction for the detection loophole introduced in \cite{MP2003}. Valid for any number of outputs, the construction is simple yet elegant. To emphasise the idea that Alice and Bob's devices are working against them, we introduce Alexa and Boris as the names of their devices, whose goal is to falsify an arbitrary non-local distribution. Beforehand they may agree a strategy (using the local hidden variable $\lambda$) but cannot communicate once they have received their input choices. Between themselves, Alexa and Boris first randomly choose a leader, with bias towards Alexa $\alpha\in[0,1]$; let us suppose for this run Alexa is chosen. They then generate uniformly a prediction for Alexa's input; say $k\in\{0,\ldots m_{\mathrm{A}}-1\}$. Finally they agree on an output $a\in\{0,\ldots n_{\mathrm{A}}-1\}$ for Alexa according to her desired marginal probability $p(a|k)$. When separated, once Alexa receives her input, if they have guessed correctly she will return outcome $a$. If the input received from Alice does not match their prediction, then Alexa outputs a failed detection $F$. It is clear this occurs with probability $(m_{\mathrm{A}}-1)/m_{\mathrm{A}}$. Meanwhile, Boris receives his input and returns $b\in\{0,\ldots n_{\mathrm{B}}-1\}$ according to $p(ab|ky)/p(a|k)$ regardless. Notice that they never jointly output a failure, so in order to fully reproduce inefficient statistics they must with some probability $\beta$ agree to both output $F$, regardless of input. This strategy gives rise to the statistics:
\begin{align}
p^{\mathrm{LHV}}(ab|xy)&=(1-\beta)\left(\frac{\alpha}{m_{\mathrm{A}}}+\frac{1-\alpha}{m_{\mathrm{B}}}\right)p(ab|xy),\nonumber\\
p^{\mathrm{LHV}}(Fb|xy)&=(1-\beta)\alpha\frac{m_{\mathrm{A}}-1}{m_{\mathrm{A}}}p(b|y),\nonumber\\
p^{\mathrm{LHV}}(aF|xy)&=(1-\beta)(1-\alpha)\frac{m_{\mathrm{B}}-1}{m_{\mathrm{B}}}p(a|x),\nonumber\\
p^{\mathrm{LHV}}(FF|xy)&=\beta.\label{LHV1}
\end{align}

One can equate equations (\ref{Eta}) and (\ref{LHV1}) to find this local hidden variable (LHV) strategy can reproduce statistics up to $\eta \leq \frac{m_{\mathrm{A}}+m_{\mathrm{B}}-2}{m_{\mathrm{A}}m_{\mathrm{B}}-1}$. By comparison to results in table \ref{Table1}, one can easily check for e.g. $m_{\mathrm{A}}=m_{\mathrm{B}}=3$  this is not optimal.\\
\begin{figure*}
\begin{equation*}\label{eq:nsform}
   \begin{array}{lllllll}
     \begin{matrix}
       \qquad\qquad\ \ \   
       \overbrace{\rule{2.0cm}{0pt}}^{m_{\mathrm{B}}-2-g}&\overbrace{\rule{1.6cm}{0pt}}^{g}
     \end{matrix}
     \\
    \begin{pmatrix}
       S & S & S &\ldots & S & L & \ldots & L \\
S & A & S/A & \ldots & S/A & L & \ldots & L \\
S & S/A & S/A & \ldots & S/A & L & \ldots & L \\
\vdots & \vdots & \vdots & & \vdots &\vdots & & \vdots\\
S & S/A & S/A & \ldots & S/A & L & \ldots & L \\
K & K & K & \ldots & K & M & \ldots & M \\
\vdots & \vdots & \vdots & & \vdots &\vdots & & \vdots\\
K & K & K & \ldots & K & M & \ldots & M
\end{pmatrix}
                                      \begin{array}{l}\phantom{\begin{matrix}0\\0\end{matrix}}\\\left.\phantom{\begin{matrix}0\\0\\0\end{matrix}}\right\rbrace{m_{\mathrm{A}}-2-h}\\\left.\phantom{\begin{matrix}0\\0\\0\end{matrix}}\right\rbrace{h}
                                      \end{array}
   \end{array}
 \end{equation*}
\begin{align*}
S=&\left(\begin{array}{cc}
\vphantom{\frac{f}{f}}\frac{1}{2} & 0\\
 0 & \frac{1}{2}\vphantom{\frac{f}{f}}
 \end{array}\right) & 
A= &\left(\begin{array}{cc}
 0 & \frac{1}{2}\vphantom{\frac{f}{f}}\\
 \frac{1}{2}\vphantom{\frac{f}{f}} & 0
 \end{array}\right) &
 K= &\left(\begin{array}{cc}
 \frac{1}{2}\vphantom{\frac{f}{f}} & \frac{1}{2}\\
 0 & 0
 \end{array}\right) &
L= &\left(\begin{array}{cc}
 \frac{1}{2}\vphantom{\frac{f}{f}} & 0  \\
 \frac{1}{2}\vphantom{\frac{f}{f}} & 0
 \end{array}\right) &
 M=&\left(\begin{array}{cc}
 1 & 0  \\
 0 & 0
 \end{array}\right)
\end{align*}
\caption{The structure of an extremal non-local distribution of $\mathcal{NS}$, where $g\in\{0,1,\ldots m_{\mathrm{B}}-2\}$, $h\in\{0,1,\ldots m_{\mathrm{A}}-2\}$. If $g$ or $h$ is non-zero, then the distribution is a lower input-number extremal point with local deterministic inputs appended to it. As such, its detection threshold cannot be lower than that of the $m_{\mathrm{A}}-h, m_{\mathrm{B}}-g$ case; and is not generally optimal for $m_{\mathrm{A}}$, $m_{\mathrm{B}}$ inputs.}
\label{Table2}
\end{figure*}

\section{A New Local Hidden Variable Construction}
\subsection{The Model}
We will look to improve this strategy on extremal binary-output $\mathcal{NS}$ points, thereby bounding the threshold for the entire space. To do this, we need to understand better the extremal points themselves. Fortunately, for binary outputs a complete characterisation has been provided in \cite{JM2005}. One can see their general form in figure \ref{Table2}. They may also be expressed in the simple form 
\begin{equation}
p(ab|xy)= \begin{cases}
1/2\;\;\; \mathrm{if}\;\;\; a \oplus b = G(x,y) = \sum_{i=1}^{2^{n_y}} Q_i(x)R_i(y) \equiv \sum_{j=1}^{2^{n_x}} S_j(y)T_j(x),\\
0\hphantom{/2}\;\;\; \mathrm{otherwise}.
\end{cases}
\end{equation}
where $Q_i(x)$ are polynomials in the binary digits \footnote{$\oplus$ denotes addition modulo 2.} of $x$, which we label $\mathbf{x}_2$, and $R_i(y)$ are  monomials in the binary digits of $y$ (labelled $\mathbf{y}_2$). Similarly, $S_j(y)$ are polynomials of $\mathbf{y}_2$ and $T_j(x)$ monomials. $n_x=\lceil \log_2 m_{\mathrm{A}} \rceil$ is the length of $\mathbf{x}_2$ and similarly for $n_y$. The most famous example of this is the (generalised) PR box \cite{PR1994}, which has the form 
\begin{equation}
p(ab|xy)=\begin{cases}
1/2 \;\;\;\mathrm{if} \;\;\; a \oplus b = \mathbf{x}_2\cdot \mathbf{y}_2 \mod 2,\\
0\hphantom{/2} \;\;\;\mathrm{otherwise}.\\
\end{cases}
\end{equation}
For all the numerically evaluated cases presented in table \ref{Table1}, the generalised PR box achieves the no-signalling threshold $\eta^*$.\\
In particular, given any extremal $\mathcal{NS}$ point, the conditional output distribution for two input pairs either match exactly or are exactly anti-matching. This allows the following strategy:
Alexa and Boris with probability $\alpha$ randomly choose a leader; suppose it is Alexa. They generate uniformly a prediction for Alexa's input; say $k_1\in\{0\ldots m_{\mathrm{A}}-1\}$; then another from the remaining $m_{\mathrm{A}}-1$ choices a second prediction, $k_2\in\{0\ldots m_{\mathrm{A}}-1\}\backslash \{k_1\}$. They also with probability $1/2$ decide whether they will use a matching or unmatching strategy. Finally, they decide uniformly on a value for $a$, $a_{\mathrm{L}}\in\{0,1\}$. Once Alexa receives her input, if it matches $k_1$ she returns outcome $a_{\mathrm{L}}$. If she receives $k_2$, for the matching strategy she returns $a_{\mathrm{L}}$, and if they are following the unmatching strategy $a_{\mathrm{L}}\oplus 1$. If her input does not match $k_1$ or $k_2$, then she outputs a failed detection $F$. It is clear this occurs with probability $(m_{\mathrm{A}}-2)/m_{\mathrm{A}}$. Meanwhile, Boris receives his input $z\in\{0\ldots m_{\mathrm{B}}-1\}$ and checks if $G(k_1,z)=G(k_2,z)$. If these values match \emph{and} they chose the matching strategy he outputs $a_{\mathrm{L}} \oplus G(k_1,z)$, otherwise he outputs $F$. If the two values are unequal and they chose the unmatching strategy he outputs $a_{\mathrm{L}} \oplus G(k_1,z)$, and $F$ otherwise. They still with some probability $\beta$ agree to both output $F$, regardless of input. This gives statistics:
\begin{align}
p^{\mathrm{LHV}}(ab|xy)&=(1-\beta)\left(\alpha\frac{1}{m_{\mathrm{A}}}+(1-\alpha)\frac{1}{m_{\mathrm{B}}}\right)p(ab|xy),\nonumber\\
p^{\mathrm{LHV}}(Fb|xy)&=(1-\beta)\left(\alpha\frac{m_{\mathrm{A}}-2}{2m_{\mathrm{A}}}+(1-\alpha)\frac{1}{m_{\mathrm{B}}}\right)p(b|y),\nonumber\\
p^{\mathrm{LHV}}(aF|xy)&=(1-\beta)\left(\alpha\frac{1}{m_{\mathrm{A}}}+(1-\alpha)\frac{m_{\mathrm{B}}-2}{2m_{\mathrm{B}}}\right)p(a|x),\nonumber\\
p^{\mathrm{LHV}}(FF|xy)&=\beta+(1-\beta)\left(\alpha\frac{m_{\mathrm{A}}-2}{2m_{\mathrm{A}}}+(1-\alpha)\frac{m_{\mathrm{B}}-2}{2m_{\mathrm{B}}}\right).\label{LHV2}
\end{align}
The advantage of such a strategy becomes apparent in the final term; to achieve the joint failure rate $(1-\eta)^2$, they can devote fewer runs to deterministically outputting $FF$, since their guessing strategy will also output a joint failure some of the time; unlike the single input guessing strategy. Equating equations (\ref{Eta}) and (\ref{LHV2}) one finds one can replicate $\eta \leq 2(m_{\mathrm{A}}+m_{\mathrm{B}}-8)/(m_{\mathrm{A}}m_{\mathrm{B}}-16)$. In the case where $m_{\mathrm{A}}=m_{\mathrm{B}}=m$, one can see this simplifies \footnote{The simplified bound also holds for $m_{\mathrm{A}}=m_{\mathrm{B}}=4$, since cancellation prevents the denominator vanishing.}  to $4/(m+4)$, which matches the known no-signalling threshold in numerically evaluated cases.

\begin{figure*}
\begin{center}
\begin{tabular}{>{\columncolor{beyes}[2pt][1.9pt]}c|c>{\columncolor{beyes}[2pt][1.9pt]}cc>{\columncolor{beyes}[2pt][1.9pt]}cc>{\columncolor{beyes}[2pt][1.9pt]}c}
Random Variable & Abort Protocol & Leader (L) & $k_1$ & $k_2$ & Match & a\\
\hline
Distribution & $P(\mathrm{yes})=\beta$ & $P(\mathrm{Alexa})=\alpha$ & $\;\;\mathbb{U}(\left\{0\ldots M_{\mathrm{L}}-1\right\})\;\;$ & $\;\;\mathbb{U}(\{0\ldots M_{\mathrm{L}}-1\}\backslash\{k_1\})\;\;$ & $P(\mathrm{Match})=1/2$ & $\mathbb{U}(\{0,1\})$\\
\hline
Outcome of $\lambda$ & no & Alexa  & 2 & 4 & Match & 0\\
\end{tabular}
$\vphantom{\frac{f}{f}}$\\
$\vphantom{\frac{f}{f}}$\\
$\vphantom{\frac{f}{f}}$\\

\resizebox{0.4\textwidth}{!}{
\begin{tabular}{c|>{\columncolor{steel}[2pt][1.9pt]}c>{\columncolor{steel}[2pt][1.9pt]}ccc}
&$\mathbf{00}$ &  $\mathbf{01}$ & $\mathbf{10}$ & $\mathbf{11}$\\
\hline
$\mathbf{00}$ &  (F,0) & (F,1) & (F,F) & (F,F)\\
\rowcolor{steel}[2pt][2pt]$\mathbf{01}$ & \cellcolor{bblue}(0,0) & \cellcolor{bblue}(0,1) & (0,F) & (0,F)\\
$\mathbf{10}$ & (F,0) &(F,1) & (F,F) & (F,F) \\
\rowcolor{steel}[2pt][2pt]$\mathbf{11}$ & \cellcolor{bblue}(0,0) & \cellcolor{bblue}(0,1)& (0,F) & (0,F)\\
\end{tabular}
}
\hspace{0.1\textwidth}
{\raggedright \resizebox{0.2\textwidth}{!}{
\begin{tabular}{cc}
\multicolumn{2}{c}{{\Large \textbf{Legend}}}\\
\hline
\\
\cellcolor{bblue}$\phantom{space fill}$ & Joint distribution reproduced\\
\\
\cellcolor{steel}$\phantom{space fill}$ & One marginal reproduced\\
\\
\cellcolor{white}$\fbox{\phantom{space fill}}$ & Both parties abort protocol\\
\end{tabular}
}}

\caption{An illustration of the improved local hidden-variable strategy. The table gives the distribution for each component variable of the underlying randomness $\lambda$. The matrix highlights for which input pairs given to Alexa and Boris they can reproduce the output correlations successfully, and when they must abort, for an example $\lambda$. The detection threshold for which one can reproduce all correlations is found by averaging over all possible $\lambda$ outcomes.}
\end{center}
\end{figure*}

\subsection{Asymptotic Power of the Model}
We now prove that the no-signalling detection threshold cannot be improved by increasing asymmetrically one party's possible measurements beyond the limit $m_{\mathrm{B}}=2^{\lceil \log_2 m_{\mathrm{A}}\rceil}$. One may express any extremal point as having $p(ab|xy)=1/2$ when $a\oplus b= G(x,y)=\sum_{j=1}^{2^{n_x}} S_j(y)T_j(x)$, with $n_x=\lceil \log_2 m_{\mathrm{A}}\rceil$. In particular this implies there are at \emph{most} $2^{n_x}$ functions of $x$ defined by the inputs of Bob. Equivalently, it implies that for any extremal point of a scenario with $m_{\mathrm{B}} > 2^{n_x}$, then for any input choice $y>2^{n_x}$ the joint distribution $p(ab|xy)$ is identical to the joint distribution $p(ab|xy')$ of some $y'\leq 2^{n_x},\, \forall\, a,b,x$. Therefore, if one has a valid LHV strategy for $m_{\mathrm{A}}, m_{\mathrm{B}}=2^{\lceil \log_2 m_{\mathrm{A}}\rceil}$ inputs up to efficiency $\eta$; one also has a valid strategy for all $m_{\mathrm{A}}, m_{\mathrm{B}} >2^{\lceil \log_2 m_{\mathrm{A}}\rceil}$ which will also achieve efficiency $\eta$. This strategy simply treats $y >2^{\lceil \log_2 m_{\mathrm{A}}\rceil}$ identically to the corresponding $y' \leq 2^{\lceil \log_2 m_{\mathrm{A}}\rceil}$.\\

\subsection{Comparison to Numerically Known No-Signalling Thresholds}
Although the bound derived in the previous section holds for all pairs $(m_{\mathrm{A}},m_{\mathrm{B}})$, we see from the numerical evidence in table \ref{Table1} it is not generally tight. In the case where $m_{\mathrm{A}}=3$, $m_{\mathrm{B}}=4$, we know the no-signalling detection threshold to be $\eta^*=5/9$; however, the hidden variable strategy we have proposed only simulates arbitrary distributions up to $\eta=1/2$. To reproduce correlations up to $\eta^*$, one can \emph{mix} our strategy with the pre-existing one \cite{MP2003} presented earlier in this paper. By choosing the pre-existing strategy, which guesses a single input, $20\%$ of the time and our strategy, predicting two inputs, $80\%$ of the time, and by choosing Alexa solely as the leader for both strategies one can achieve $\eta \leq 5/9$. This mixing of strategies does not extend to higher dimensional asymmetric scenarios though; for $m_{\mathrm{A}}=5,\;m_{\mathrm{B}}=6$ no combination of the two strategies beats the bound given by equation (\ref{LHV2}).


As the number of input choices increases, one could propose a more general variation; in which the leader (say Alexa) chooses many input predictions $k_1\ldots k_n \in \{0\ldots m_{\mathrm{A}}-1\}$, $n\leq m_{\mathrm{A}}$. With this strategy, they must beforehand predict whether $G(z_i,k)$ will coincide with $G(z_1,k)$, for each $i=2\ldots n$. This is analogous to the `matching/unmatching' choice seen earlier. The probability of guessing this correctly scales as $2^{n-1}$. However, the benefit of predicting additional inputs only scales as $n/m_{A}$. This implies the probability of a correct output will scale as $\frac{n}{m_{\mathrm{A}}}\frac{1}{2^{n-1}}$, which takes its maximal value at $n=1,2$ only. Trying to incorporate this strategy to simulate $m_{\mathrm{A}}=5,\;m_{\mathrm{B}}=6$ distributions, our optimisation never chose strategies with $n>2$. This suggests for the asymmetric case a more nuanced joint strategy is required. However, we stress that when $m_{\mathrm{A}}=m_{\mathrm{B}}$, the bound predicted by this model matches all numerically obtained bounds.\\


In order to prove that our conjecture of $\eta^*=4/(m+4)$ for $m_{\mathrm{A}}=m_{\mathrm{B}}=m$ is correct, one would need to provide a extremal $\mathcal{NS}$ distribution $p^{\mathrm{NS}}$, and corresponding Bell inequality $s_{a'b'}^{xy}$, such that $\sum_{a',b',x,y}s_{a'b'}^{xy}p^{\mathrm{NS}}_{\eta}(a'b'|xy) \not \leq k,\; \forall \eta > \eta^*$. Here we have used $a',\,b'$ to explicitly remind the reader that $a'$ ranges both in the original values of $a$ and $F$; that is, it is a 3-outcome inequality. From numerical results, the generalised PR box is the best candidate for the extremal point, but we found no obvious generalisation of the witnessing Bell inequalities, which are provided for evaluated cases in the appendix.\\

\subsection{Comparison to Quantumly Realisable Thresholds}
As stated above, to prove the no-signalling threshold $\eta^*$ for a given scenario requires a Bell inequality violation $\sum_{a',b',x,y}s_{a'b'}^{xy}p^{\mathrm{NS}}_{\eta}(a'b'|xy) \not \leq k$, $\forall \eta>\eta^*$. Therefore $s_{a'b'}^{xy}$ is the `optimal' Bell inequality, in that it detects non-locality for all efficiencies above the no-signalling threshold. A natural question is whether the same Bell inequality is optimal with respect to quantum correlations; i.e. $\sum_{a',b',x,y}s_{a'b'}^{xy}p^{\mathrm{Q}}_{\eta}(a'b'|xy) \not \leq k$, $\forall \eta>\eta_c$.

 For quantum correlations, a critical efficiency of $\eta_c=2/3$ is achievable via qubits using the $m_{\mathrm{A}}=m_{\mathrm{B}}=2$ CHSH inequality \cite{E1993}, whilst testing ququarts with a $m_{\mathrm{A}}=m_{\mathrm{B}}=4$ inequality allows a  critical efficiency of $(\sqrt{5}-1)/2\approx 0.618$. The respective Bell inequalities verifying non-locality for efficiencies higher than the critical efficiency, when applied to the generalised PR box achieve the no-signalling detection threshold, $\eta^*$, for their respective scenarios. These inequalities are somewhat special in that they are `lifted inequalities'; they are of the form
\begin{align*}
	s_{\mathrm{F}b|xy}&=s_{a_{x}b|xy}, & s_{a\mathrm{F}|xy}&=s_{ab_{y}|xy},\; a_x,b_y\in\{0,1\}\; \forall x,y
\end{align*}
i.e. facet 2-outcome inequalities where $F$ is treated identically to one of the valid outputs.
In contrast, the optimal Bell inequality for $m_{\mathrm{A}}=m_{\mathrm{B}}=3$ requires a truly new 3-output inequality; something noted in \cite{WDAP2008}.

In order to test whether our optimal Bell inequalities could lead to new quantum constructions, we employed the NPA hierarchy of correlations \cite{NPA2008}. These allow one to define successively tighter outer approximations to $\mathcal{Q}$, which we label $\mathcal{Q}_1\supset \mathcal{Q}_2\ldots \supset Q$. For a fixed $\eta$, we can then employ semidefinite programming to look for a set of correlations such that $\mathbf{p}\in \mathcal{Q}_{i}, \mathbf{s}\cdot\mathbf{p}_{\eta}\not\leq k,$, which implies $\mathbf{p}_{\eta} \not\in \mathcal{L}$. It is then clear that, if for a given $i$, $\tilde{\eta}$  no such  $\mathbf{p}$ is found, then $\{\mathbf{p}\in \mathcal{Q}, \mathbf{s}\cdot\mathbf{p}_{\tilde{\eta}}\not\leq k \}$ must also be empty. \\

For the scenarios $m_{\mathrm{A}}=m_{\mathrm{B}}=3$ and $m_{\mathrm{A}}=3, m_{\mathrm{B}}=4$, we know the quantum critical efficiency is not higher than $2/3$; since we may always embed the CHSH /qubit construction into these scenarios. Therefore, an improvement in the quantum critical efficiency would require that $\{\mathbf{p}\in \mathcal{Q}, \mathbf{s}\cdot\mathbf{p}_{2/3}\not\leq k \}$ is non-empty. However, in both scenarios, choosing $\mathbf{s}$ as the optimal Bell inequality for non-locality, we find that this set is empty at level $\mathcal{Q}_2$ of the hierarchy; thus these inequalities do not help us to improve the quantum critical efficiency, $\eta_c$.

\section{Conclusions and Discussion}
In this paper, we have exploited the structure of the bipartite binary-output no-signalling polytope in order to provide a lower bound on the detection loophole critical efficiency for an arbitrary number of inputs. We have done this by constructing an explicit local hidden-variable model valid for all extremal points. Numerical evidence suggests that when Alice and Bob share an equal number of inputs, this construction is optimal. An open question is whether one can find a family of Bell inequalities verifying this.\\

One possible extension to this work would be improve the strategy for asymmetric measurement capabilities; since we know our model does not provide a tight bound for $m_{\mathrm{A}}=5$, $m_{\mathrm{B}}=6$. A further generalisation would be to test if this approach generalises to a larger number of outputs. Unfortunately, the vertices of higher output no-signalling polytopes are not generally known, so we cannot say much about their structure. Considering the results here, one would expect the successful simulation efficiency of a construction which predicts $n$ inputs in a $k$-output scenario to scale as $\frac{n}{m_{\mathrm{A}}}\frac{1}{k^{n-1}}$, which for $k>2$ achieves optimal integer value only at $n=1$. This suggests for higher output-number scenarios the construction of \cite{MP2003}, defining equation (\ref{LHV1}), may be optimal.\\
\section{Acknowledgements}
This work was supported, in part, by the DFG through SFB 1227 (DQ-mat), the RTG 1991, and funded by the Deutsche Forschungsgemeinschaft (DFG, German Research Foundation) under Germany’s Excellence Strategy – EXC-2123 Quantum Frontiers – 390837967. We would like to thank Tobias Osborne, Reinhard Werner and Le Phuc Thinh for useful discussions.

\bibliographystyle{naturemag}

\newpage
\onecolumngrid
\appendix

\section{Local Weight Linear Program}
In order to calculate the linear weight of an arbitrary distribution $\mathbf{q}$, we solve the following problem:\\
\begin{center}
Maximise $\sum_{i} \alpha_i$, subject to: $\sum_{i} \alpha_i \mathbf{q}_{i}^\mathcal{L} \leq \mathbf{q}$, $\alpha_i \geq 0$.\\
\end{center}
where $\mathbf{q}_{i}^\mathcal{L}$ are the extremal points of the polytope $\mathcal{L}$. By rearranging the inequality, we see that the leftover distribution $\mathbf{q}':=\mathbf{q}-\sum_{i} \alpha \mathbf{q}_{i}^\mathcal{L}$ has all positive entries, and satisfies the no-signalling constraints since so too do $\mathbf{q},\mathbf{q}_{i}^\mathcal{L}$. Therefore it is a valid (sub-normalised) distribution. This linear program therefore looks to optimise the total weight of the local extremal points over all decompositions of $\mathbf{q}$.\\

It is also worth mentioning that every linear program has a dual with the same optimal value \cite{BV2004}. The dual of the above function gives us a vector $\mathbf{b}$ such that:\\
\begin{center}
$\mathbf{b}^{\mathrm{T}}\mathbf{q} = \sum_{i} \alpha_i$, $\mathbf{b}^{\mathrm{T}}\mathbf{q}_{i}^\mathcal{L} \geq 1 \forall i$. \\
\end{center}
we see immediately that if $\sum_{i} \alpha_i < 1$, this gives us a Bell inequality violated by $\mathbf{q}$.

\section{Bell Inequalities which verify the Threshold}
In this supplemental file, the optimal Bell inequalities are provided to achieve the detection loophole threshold for the generalised PR box. They are presented in matrix format:

\begin{equation}\label{tableform}
S=\left(\begin{array}{ccc|cccc|ccc}
s_{0,0}^{0,0} & \ldots & s_{0,n_B}^{0,0} & \ldots&\ldots&\ldots&\ldots & s_{0,0}^{0,m_B-1} & \ldots & s_{0,n_B}^{0,m_B-1} \\
\vdots & \ddots & \vdots & \ldots&\ldots&\ldots&\ldots & \vdots & \ddots & \vdots \\
s_{n_A,0}^{0,0} & \ldots & s_{n_A,n_B}^{0,0} & \ldots&\ldots&\ldots&\ldots & s_{n_A,0}^{0,m_B-1} & \ldots & s_{n_A,n_B}^{0,m_B-1}\\
\hline
\vdots & \vdots & \vdots & \ddots &&& & \vdots & \vdots &\vdots\\
\vdots & \vdots & \vdots & &&& & \vdots & \vdots &\vdots\\
\hline
s_{0,0}^{m_A-1,0} & \ldots & s_{0,n_B}^{m_A-1,0} & \ldots&\ldots&\ldots&\ldots & s_{0,0}^{m_A-1,m_B-1} & \ldots & s_{0,n_B}^{m_A-1,m_B-1} \\
\vdots & \ddots & \vdots & \ldots&\ldots&\ldots&\ldots & \vdots & \ddots & \vdots \\
s_{n_A,0}^{m_A-1,0} & \ldots & s_{n_A,n_B}^{m_A-1,0} & \ldots&\ldots&\ldots&\ldots &s_{n_A,0}^{m_A-1,m_B-1}&\ldots &s_{n_A,n_B}^{m_A-1,m_B-1}
\end{array}\right),
\end{equation}
where the solid lines delineate different inputs. All presented inequalities have local bound $\geq 1$. Note that there are $n_A+1$ ($n_B+1$) outputs to account for the additional output $F$.\\

\subsection{Optimal Inequality for 2-Inputs}
As mentioned in the main body of the paper, this inequality is a `lifting' of the CHSH inequality. For all measurements failure to output is treated identically to $0$. Since other liftings of the same CHSH inequality achieve the optimal value, we can see generally there is not a single unique inequality that witnesses the threshold. 
\begin{equation}
\left(
\begin{array}{ccc|ccc}
 0 & 1 & 0 & 0 & 1 & 0 \\
 1 & 0 & 1 & 1 & 0 & 1 \\
 0 & 1 & 0 & 0 & 1 & 0 \\
 \hline
 0 & 1 & 0 & 1 & 0 & 1 \\
 1 & 0 & 1 & 0 & 1 & 0 \\
 0 & 1 & 0 & 1 & 0 & 1 \\
\end{array}
\right)
\end{equation}
\subsection{Optimal Inequality for 3-Inputs}
Unlike the previous case; this inequality is a `true' 3-input, 3-output inequality; it cannot be created from lifting a previous, lower dimensional inequality. What is interesting to note is that, for the first two inputs for each party, failure is again treated identically to $0$ --- it is only the final input which treats failure differently.
\begin{equation}
\left(
\begin{array}{ccc|ccc|ccc}
 0 & \vphantom{\frac{f}{f}} \frac{1}{3} & 0 & 0 & \vphantom{\frac{f}{f}} \frac{1}{3} & 0 & 0 & \vphantom{\frac{f}{f}} \frac{2}{3} & 0 \\
 \vphantom{\frac{f}{f}} \frac{1}{3} & 0 & \vphantom{\frac{f}{f}} \frac{1}{3} & \vphantom{\frac{f}{f}} \frac{1}{3} & 0 & \vphantom{\frac{f}{f}} \frac{1}{3} & \vphantom{\frac{f}{f}} \frac{2}{3} & 0 & 0 \\
 0 & \vphantom{\frac{f}{f}} \frac{1}{3} & 0 & 0 & \vphantom{\frac{f}{f}} \frac{1}{3} & 0 & 0 & \vphantom{\frac{f}{f}} \frac{2}{3} & 0 \\
  \hline
 0 & \vphantom{\frac{f}{f}} \frac{1}{3} & 0 & \vphantom{\frac{f}{f}} \frac{1}{3} & 0 & \vphantom{\frac{f}{f}} \frac{1}{3} & 0 & 0 & 0 \\
 \vphantom{\frac{f}{f}} \frac{1}{3} & 0 & \vphantom{\frac{f}{f}} \frac{1}{3} & 0 & \vphantom{\frac{f}{f}} \frac{1}{3} & 0 & \frac{2}{3} & 0 & 0 \\
 0 & \vphantom{\frac{f}{f}} \frac{1}{3} & 0 & \vphantom{\frac{f}{f}} \frac{1}{3} & 0 & \vphantom{\frac{f}{f}} \frac{1}{3} & 0 & 0 & 0 \\
  \hline
 0 & \vphantom{\frac{f}{f}} \frac{2}{3} & 0 & 0 & \frac{2}{3} & 0 & \vphantom{\frac{f}{f}} \frac{2}{3} & 0 & \vphantom{\frac{f}{f}} \frac{2}{3} \\
 \vphantom{\frac{f}{f}} \frac{2}{3} & 0 & \vphantom{\frac{f}{f}} \frac{2}{3} & 0 & 0 & 0 & 0 & \vphantom{\frac{f}{f}} \frac{2}{3} & \vphantom{\frac{f}{f}} \frac{2}{3} \\
 0 & 0 & 0 & 0 & 0 & 0 & \vphantom{\frac{f}{f}} \frac{2}{3} & \vphantom{\frac{f}{f}} \frac{2}{3} & \vphantom{\frac{f}{f}} \frac{2}{3} \\
\end{array}
\right)
\end{equation}
\subsection{Optimal Inequality for 4-Inputs}
This inequality is also a lifting of a 4-input, 2-output inequality; however in this instance the choice of treating failure as $0$ or $1$ depends on the input.
\begin{equation}
\left(
\begin{array}{ccc|ccc|ccc|ccc}
 0 & \vphantom{\frac{f}{f}} \frac{1}{4} & 0 & 0 & \vphantom{\frac{f}{f}} \frac{1}{4} & 0 & 0 & \vphantom{\frac{f}{f}} \frac{1}{4} & \vphantom{\frac{f}{f}} \frac{1}{4} & 0 & \vphantom{\frac{f}{f}} \frac{1}{4} & 0 \\
 \vphantom{\frac{f}{f}} \frac{1}{4} & 0 & \vphantom{\frac{f}{f}} \frac{1}{4} & \vphantom{\frac{f}{f}} \frac{1}{4} & 0 & \vphantom{\frac{f}{f}} \frac{1}{4} & \vphantom{\frac{f}{f}} \frac{1}{4} & 0 & 0 & \vphantom{\frac{f}{f}} \frac{1}{4} & 0 & \vphantom{\frac{f}{f}} \frac{1}{4} \\
 0 & \vphantom{\frac{f}{f}} \frac{1}{4} & 0 & 0 & \vphantom{\frac{f}{f}} \frac{1}{4} & 0 & 0 & \vphantom{\frac{f}{f}} \frac{1}{4} & \vphantom{\frac{f}{f}} \frac{1}{4} & 0 & \vphantom{\frac{f}{f}} \frac{1}{4} & 0 \\
  \hline
 0 & \vphantom{\frac{f}{f}} \frac{1}{4} & 0 & \vphantom{\frac{f}{f}} \frac{1}{4} & 0 & \vphantom{\frac{f}{f}} \frac{1}{4} & 0 & 0 & 0 & 0 & 0 & 0 \\
 \vphantom{\frac{f}{f}} \frac{1}{4} & 0 & \vphantom{\frac{f}{f}} \frac{1}{4} & 0 & \vphantom{\frac{f}{f}} \frac{1}{4} & 0 & \vphantom{\frac{f}{f}} \frac{1}{2} & 0 & 0 & 0 & \vphantom{\frac{f}{f}} \frac{1}{2} & 0 \\
 0 & \vphantom{\frac{f}{f}} \frac{1}{4} & 0 & \vphantom{\frac{f}{f}} \frac{1}{4} & 0 & \vphantom{\frac{f}{f}} \frac{1}{4} & 0 & 0 & 0 & 0 & 0 & 0 \\
  \hline
 0 & \vphantom{\frac{f}{f}} \frac{1}{4} & 0 & 0 & \vphantom{\frac{f}{f}} \frac{1}{2} & 0 & \vphantom{\frac{f}{f}} \frac{1}{4} & 0 & 0 & \vphantom{\frac{f}{f}} \frac{1}{2} & 0 & \vphantom{\frac{f}{f}} \frac{1}{2} \\
 \vphantom{\frac{f}{f}} \frac{1}{4} & 0 & \vphantom{\frac{f}{f}} \frac{1}{4} & 0 & 0 & 0 & 0 & \vphantom{\frac{f}{f}} \frac{1}{4} & \vphantom{\frac{f}{f}} \frac{1}{4} & 0 & 0 & 0 \\
 \vphantom{\frac{f}{f}} \frac{1}{4} & 0 & \vphantom{\frac{f}{f}} \frac{1}{4} & 0 & 0 & 0 & 0 & \vphantom{\frac{f}{f}} \frac{1}{4} & \vphantom{\frac{f}{f}} \frac{1}{4} & 0 & 0 & 0 \\
  \hline
 0 & \vphantom{\frac{f}{f}} \frac{1}{4} & 0 & 0 & 0 & 0 & \vphantom{\frac{f}{f}} \frac{1}{2} & 0 & 0 & 0 & \vphantom{\frac{f}{f}} \frac{1}{4} & 0 \\
 \vphantom{\frac{f}{f}} \frac{1}{4} & 0 & \vphantom{\frac{f}{f}} \frac{1}{4} & 0 & \vphantom{\frac{f}{f}} \frac{1}{2} & 0 & 0 & 0 & 0 & \vphantom{\frac{f}{f}} \frac{1}{4} & 0 & \vphantom{\frac{f}{f}} \frac{1}{4} \\
 0 & \vphantom{\frac{f}{f}} \frac{1}{4} & 0 & 0 & 0 & 0 & \vphantom{\frac{f}{f}} \frac{1}{2} & 0 & 0 & 0 & \vphantom{\frac{f}{f}} \frac{1}{4} & 0 \\
\end{array}
\right)
\end{equation}

\subsection{Optimal Inequality for an Asymmetric Case: Alice 3 Inputs, Bob 4 Inputs}
For this asymmetric case we can again provide a Bell inequality which achieves the optimal threshold for the generalised PR Box. Like the previous cases, the inequality we provide here is a \emph{facet} inequality; that is a maximally dimensional face of the local polytope. This is the first inequality provided where the failure outcome is treated differently from the valid outcomes for \emph{all} input choices; we leave open the question whether this is necessary, or an artefact of this particular inequality.
\begin{equation}\left(
\begin{array}{ccc|ccc|ccc|ccc}
 0 & \vphantom{\frac{f}{f}} \frac{1}{3} & \vphantom{\frac{f}{f}} \frac{1}{6} & 0 & \vphantom{\frac{f}{f}} \frac{1}{3} & \vphantom{\frac{f}{f}} \frac{1}{6} & 0 & \vphantom{\frac{f}{f}} \frac{1}{3} & \vphantom{\frac{f}{f}} \frac{1}{4} & 0 & \vphantom{\frac{f}{f}} \frac{1}{4} & 0 \\
 \vphantom{\frac{f}{f}} \frac{5}{12} & 0 & \vphantom{\frac{f}{f}} \frac{1}{4} & \vphantom{\frac{f}{f}} \frac{1}{3} & 0 & \vphantom{\frac{f}{f}} \frac{1}{6} & \vphantom{\frac{f}{f}} \frac{1}{3} & 0 & \vphantom{\frac{f}{f}} \frac{1}{12} & \vphantom{\frac{f}{f}} \frac{1}{2} & 0 & \vphantom{\frac{f}{f}} \frac{1}{4} \\
 \vphantom{\frac{f}{f}} \frac{1}{2} & \vphantom{\frac{f}{f}} \frac{5}{12} & \vphantom{\frac{f}{f}} \frac{5}{12} & 0 & 0 & 0 & 0 & 0 & 0 & \vphantom{\frac{f}{f}} \frac{1}{12} & 0 & 0 \\
 \hline
 0 & \vphantom{\frac{f}{f}} \frac{1}{3} & \vphantom{\frac{f}{f}} \frac{1}{12} & 0 & 0 & 0 & 0 & \vphantom{\frac{f}{f}} \frac{7}{12} & \vphantom{\frac{f}{f}} \frac{1}{3} & \vphantom{\frac{f}{f}} \frac{1}{3} & 0 & \vphantom{\frac{f}{f}} \frac{1}{6} \\
 \vphantom{\frac{f}{f}} \frac{1}{4} & 0 & 0 & 0 & \vphantom{\frac{f}{f}} \frac{1}{2} & \vphantom{\frac{f}{f}} \frac{1}{6} & 0 & 0 & 0 & 0 & \vphantom{\frac{f}{f}} \frac{1}{2} & \vphantom{\frac{f}{f}} \frac{1}{12} \\
 0 & \vphantom{\frac{f}{f}} \frac{1}{6} & 0 & 0 & \vphantom{\frac{f}{f}} \frac{1}{6} & 0 & \vphantom{\frac{f}{f}} \frac{1}{6} & \vphantom{\frac{f}{f}} \frac{1}{4} & \vphantom{\frac{f}{f}} \frac{1}{6} & 0 & \vphantom{\frac{f}{f}} \frac{1}{4} & 0 \\
   \hline
 0 & \vphantom{\frac{f}{f}} \frac{1}{6} & \vphantom{\frac{f}{f}} \frac{1}{12} & 0 & 0 & 0 & \vphantom{\frac{f}{f}} \frac{3}{4} & 0 & \vphantom{\frac{f}{f}} \frac{1}{6} & \vphantom{\frac{f}{f}} \frac{5}{12} & 0 & \vphantom{\frac{f}{f}} \frac{5}{12} \\
\vphantom{\frac{f}{f}} \frac{1}{3} & 0 & \vphantom{\frac{f}{f}} \frac{1}{6} & \vphantom{\frac{f}{f}} \frac{2}{3} & 0 & \vphantom{\frac{f}{f}} \frac{1}{3} & 0 & 0 & 0 & 0 & \vphantom{\frac{f}{f}} \frac{1}{3} & \vphantom{\frac{f}{f}} \frac{1}{6} \\
 0 & 0 & 0 & \vphantom{\frac{f}{f}} \frac{1}{3} & 0 & \vphantom{\frac{f}{f}} \frac{1}{6} & \vphantom{\frac{f}{f}} \frac{1}{2} & \vphantom{\frac{f}{f}} \frac{1}{6} & \vphantom{\frac{f}{f}} \frac{1}{4} & 0 & 0 & 0 \\
\end{array}
\right)
\end{equation}

\subsection{Optimal Inequality for 5-Inputs}
The previous inequalities provided were all calculated using exact arithmetic. Unfortunately this takes much longer than floating point methods, particularly as the dimension increases. Therefore, we are only able to provide a Bell inequality here which is accurate up to 6 s.f. and moreover, not a facet inequality. However, it still verifies the detection loophole threshold, and is included for completeness.
\begin{equation}
\left(
\begin{array}{cccccccccc}
	0 & 144.708 & 0.0476284 & 0 & 147.002 & 0.0560145 & 0 & 147.002 & 0.0560145 & 0 \\
	147.002 & 0.0560145 & 0 & 148.982 & 0.0504226 & 144.708 & 0 & 0.0476284 & 147.002 & 0 \\
	0.0560145 & 147.002 & 0 & 0.0560145 & 147.002 & 0 & 0.0560145 & 148.982 & 0 & 0.0504226 \\
	0.0476289 & 0.0476289 & 0.100928 & 0.0494062 & 0.0494062 & 0.0326552 & 0.0494062 & 0.0494062 & 0.0326552 & 0.0494062 \\
	0.0494062 & 0.0326552 & 0.0712469 & 0.0712471 & 0.000105974 & 0 & 147.002 & 0.0494056 & 147.860 & 0 \\
	0.0582717 & 0 & 149.205 & 0.0510420 & 147.860 & 0 & 0.0582717 & 0 & 145.578 & 0.0573425 \\
	147.002 & 0 & 0.0494056 & 0 & 147.860 & 0.0582717 & 149.205 & 0 & 0.0510421 & 0 \\
	147.860 & 0.0582717 & 145.578 & 0 & 0.0573426 & 0.0560145 & 0.0560145 & 0.0326547 & 0.0582717 & 0.0582717 \\
	0.0336107 & 0.0510419 & 0.0510421 & 0.0221489 & 0.0582717 & 0.0582717 & 0.0336107 & 0.0578340 & 0.0578340 & 0.0708747 \\
	0 & 147.002 & 0.0494056 & 0 & 149.205 & 0.0510420 & 147.860 & 0 & 0.0582717 & 147.860 \\
	0 & 0.0582717 & 0 & 145.578 & 0.0573425 & 147.002 & 0 & 0.0494056 & 149.205 & 0 \\
	0.0510421 & 0 & 147.860 & 0.0582717 & 0 & 147.860 & 0.0582717 & 145.578 & 0 & 0.0573426 \\
	0.0560145 & 0.0560146 & 0.0326547 & 0.0510419 & 0.0510421 & 0.0221489 & 0.0582717 & 0.0582717 & 0.0336107 & 0.0582717 \\
	0.0582717 & 0.0336107 & 0.0578340 & 0.0578340 & 0.0708747 & 0 & 147.002 & 0.0494056 & 147.860 & 0 \\
	0.0582717 & 147.860 & 0 & 0.0582717 & 0 & 149.205 & 0.0510420 & 0 & 145.578 & 0.0573425 \\
	147.002 & 0 & 0.0494056 & 0 & 147.860 & 0.0582717 & 0 & 147.860 & 0.0582717 & 149.205 \\
	0 & 0.0510421 & 145.578 & 0 & 0.0573426 & 0.0560145 & 0.0560146 & 0.0326547 & 0.0582717 & 0.0582717 \\
	0.0336107 & 0.0582717 & 0.0582717 & 0.0336107 & 0.0510419 & 0.0510421 & 0.0221489 & 0.0578340 & 0.0578340 & 0.0708747 \\
	0 & 148.982 & 0.0712492 & 0 & 145.578 & 0.0578340 & 0 & 145.578 & 0.0578340 & 0 \\
	145.578 & 0.0578340 & 149.070 & 0 & 0.0661534 & 148.982 & 0 & 0.0712493 & 145.578 & 0 \\
	0.0578341 & 145.578 & 0 & 0.0578341 & 145.578 & 0 & 0.0578341 & 0 & 149.070 & 0.0661534 \\
	0.0504204 & 0.0504204 & 0.000106005 & 0.0573429 & 0.0573430 & 0.0708752 & 0.0573429 & 0.0573430 & 0.0708752 & 0.0573429 \\
\end{array}
\right)
\end{equation}

\end{document}